\documentclass[prl,superscriptaddress,twocolumn,floatfix]{revtex4}
\usepackage{amssymb}
\usepackage{amsmath}
\usepackage{graphicx}
\usepackage{color}

\newcommand{\by}{$B_{y}$}

\begin{document}

\title{Superconductor-Nanowire Devices from Tunneling to the Multichannel Regime: \\ Zero-Bias Oscillations and Magnetoconductance Crossover}
\author{H.~O.~H.~Churchill}
\affiliation{Department of Physics, Harvard University, Cambridge, Massachusetts 02138, USA}
\affiliation{Department of Physics, Massachusetts Institute of Technology, Cambridge, Massachusetts 02139, USA}
\author{V.~Fatemi}
\affiliation{Department of Physics, Massachusetts Institute of Technology, Cambridge, Massachusetts 02139, USA}
\author{K.~Grove-Rasmussen}
\affiliation{Center for Quantum Devices, Niels Bohr Institute, University of Copenhagen, 2100 Copenhagen \O, Denmark}
\author{M.~T.~Deng}
\affiliation{Division of Solid State Physics, Lund University, Box 118, S-221 00 Lund, Sweden}
\author{P.~Caroff}
\affiliation{Division of Solid State Physics, Lund University, Box 118, S-221 00 Lund, Sweden}
\author{H.~Q.~Xu}
\affiliation{Division of Solid State Physics, Lund University, Box 118, S-221 00 Lund, Sweden}
\affiliation{Department of Electronics and Key Laboratory for the Physics and Chemistry of Nanodevices, Peking University, Beijing 100871, China}
\author{C.~M.~Marcus}
\email{marcus@nbi.dk}
\affiliation{Center for Quantum Devices, Niels Bohr Institute, University of Copenhagen, 2100 Copenhagen \O, Denmark}

\begin{abstract}
We present transport measurements in superconductor-nanowire devices with a gated constriction forming a quantum point contact.
Zero-bias features in tunneling spectroscopy appear at finite magnetic fields, and oscillate in amplitude and split away from zero bias as a function of magnetic field and gate voltage.
A crossover in magnetoconductance is observed:  Magnetic fields above $\sim$0.5~T enhance conductance in the low-conductance (tunneling) regime but suppress conductance in the high-conductance (multichannel) regime. We consider these results in the context of Majorana zero modes as well as alternatives, including Kondo effect and analogs of 0.7 structure in a disordered nanowire.
\end{abstract}

\maketitle
Physical systems with Majorana quasiparticles---zero-energy modes with non-Abelian exchange statistics---represent a topological phase of matter that could form the basis of topologically-protected quantum computation \cite{Kitaev:2003ul,Nayak:2008wx,Wilczek:2009wv}.
Pursuit of such systems has been advanced by a range of proposals including $\nu=5/2$ fractional quatum Hall states \cite{Moore:1991vr}, $p$-wave superconductors \cite{Read:2000vz}, cold atom systems \cite{Gurarie:2005vg,Jiang:2011wm}, and hybrid systems of $s$-wave superconductors with either topological insulators \cite{Fu:2008gu} or semiconductors \cite{Kitaev:2001up,Sau:2010cl,Alicea:2010te}.
An attractive implementation calls for coupling an s-wave superconductor to a one-dimensional semiconductor nanowire with strong spin-orbit coupling.  In a magnetic field, tuning the chemical potential of the nanowire so that the induced superconducting gap lies well within the Zeeman splitting permits effective p-wave superconductivity supporting Majorana end-state zero modes \cite{Oreg:2010gk,Lutchyn:2010hp}.

Expected signatures of a topological phase in the nanowire system include a zero-bias conductance peak \cite{Law:2009gf,Flensberg:2010uw,Sau:2010gh} and fractional Josephson effect \cite{Kitaev:2001up}, both of which have been reported as evidence of Majorana fermions \cite{Mourik:2012wk,Das:2012th,Deng:2012vp,RokhinsonLP:2012wy}.
The peak is predicted to oscillate about zero energy as a function of magnetic field and chemical potential \cite{Prada:2012iv,Rainis:2012wn,DasSarma:2012kt}. 
Features suggesting this effect have been reported in Ref.~\onlinecite{Finck:2012tc} and considered both in the context of Majorana zero modes and the Kondo effect.
Given the interest in realizing topological states of matter and non-Abelian quasiparticle statistics, it is imperative to broaden the range of experimental observations and to consider interpretations in the context of Majorana modes as well as alternatives such as the Kondo effect \cite{DGG:2012ut,Nilsson:2009vf,Lee:2012uq}, 0.7 structure \cite{Thomas:1996wq,Cronenwett:2002ui,Meir:2002tw,Ren:2010vf}, weak anti-localization \cite{Pikulin:2012fs}, and disorder-induced level crossings \cite{Bagrets:2012ta,Liu:2012vj}.

Here, we report transport measurements in superconductor-nanowire devices configured as a quantum point contact (QPC) over a broad range of magnetic fields and conductances from the tunneling regime to the multichannel regime.
We deliberately tuned the device to a regime without evidence of dot-like charging features or even-odd structure (see Supplementary Material).
We observed zero-bias features in tunneling spectroscopy above $\sim$0.5~T that oscillated in amplitude and bias position as a function of magnetic field and gate voltage. 
We also observed that the zero-bias conductance of the QPC was enhanced by a magnetic field near pinch-off and suppressed at higher transmission, in qualitative agreement with the trends described in Ref.~\onlinecite{Wimmer:2011vl} for the trivial-to-topological crossover.
These results are consistent with some but not all predictions for Majorana zero modes, and do not yet rule out alternative explanations such as Kondo-enhanced conductance in confined structures or zero-bias peaks in single-barrier structures analogous to 0.7 structure in QPCs.

%:Fig 1
%-------------------%
\begin{figure}[t]
\center \label{figure1}
\includegraphics[width=3.4in]{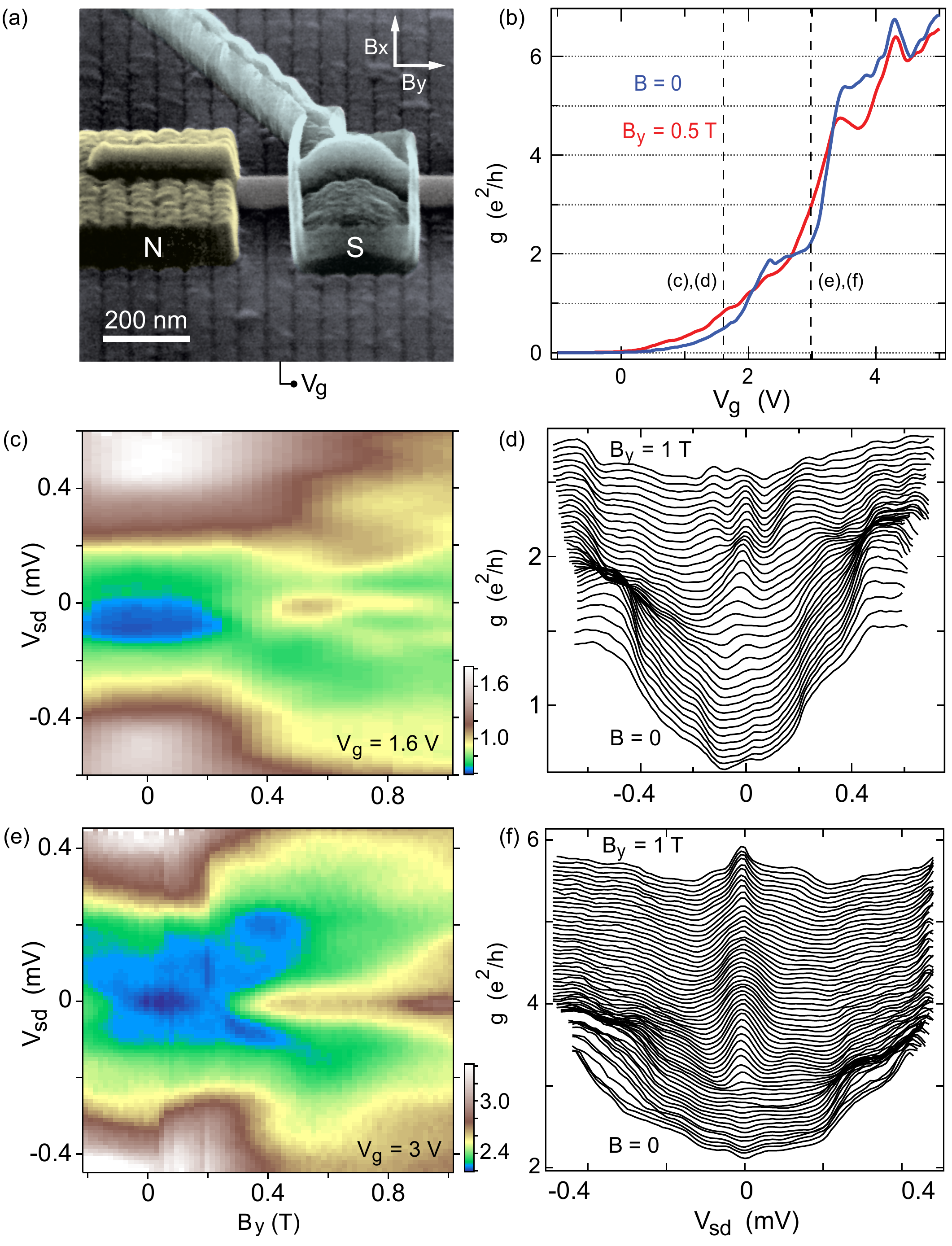}
\caption{\footnotesize{(a) Scanning electron micrograph of one of the devices measured (device 2).  An InSb nanowire was deposited over HfO$_2$-insulated bottom gates, and contacted by Ti/Au leads (N) on the ends and a Ti/NbTiN lead (S) in the middle.  (b) Zero-bias differential conductance, $g$, at ${\rm B}=0$ (blue curve) and \by$=0.5$ T (red curve) as a function of the voltage, $V_g$, on the gate labeled in (a).  (c) $g$ as a function of \by~and $V_{\rm sd}$ applied to the normal lead for $V_g=1.6$ V [left dashed line in (b)]. (d) Cuts from (c) for \by~between 0 and 1 T, offset for clarity except for the B = 0 trace.  (e) $g$ as a function of \by~and $V_{\rm sd}$ for $V_g=3$ V [right dashed line in (b)]. (f) Cuts from (e) for \by~between 0 and 1 T, offset for clarity except for the B = 0 trace.  All data from device 1.}}
\end{figure}
%-------------------%

InSb nanowires with diameter 100 nm \cite{Caroff:2008wv,Nilsson:2009vf} were contacted by a superconducting lead (1/150 nm Ti/Nb$_{0.7}$Ti$_{0.3}$N) and one or two normal leads covering the wire ends (5/125 nm Ti/Au).
Data from two devices are reported. Device 1 had normal leads on both ends, and device 2 had one normal lead, as in Fig.~1(a).
The width of the superconducting lead was 300 nm for device 1 and 250 nm for device 2, and the length of nanowire between the superconducting and normal leads was 150 nm for device 1 and 100 nm for device 2.
The coupling to the normal leads was tuned by local control of the electron density in the nanowire using bottom-gates that were insulated by 30 nm of HfO$_2$ \cite{Biercuk:2003uc}. 
Some gates were under the region of the nanowire covered by the superconductor, and some gates were under the uncovered region.
The samples were measured in a dilution refrigerator using standard lock-in techniques. 

Control of the coupling between the superconducting and normal sections of the device is demonstrated in Fig.~1(b) by a measurement of the zero-bias differential conductance, $g$, as a function of bottom-gate voltage, $V_g$, for device 1.
With the voltages on the other bottom-gates set to 3 V, $g$ varied from $7e^2/h$ at $V_g=5$ V to zero for $V_g < 0$ V \cite{Gnote}.
A plateau-like shoulder at $g\sim 2 e^2/h$ is evident at $B=0$ around $V_g=2.5$ V. 
This value of conductance is a factor of two smaller than expected for the conductance of the first plateau for a QPC in perfect contact with a superconductor \cite{Wimmer:2011vl}.
In a magnetic field \by$=0.5$~T along the wire axis, $g$ increased in two regions of gate voltage [dashed vertical lines in Fig.~1(b)], and decreased at larger conductances, $V_g>3.5$ V. 

At finite source-drain voltage, $V_{\rm sd}$, differential conductance increased steeply around $V_{\rm sd}=\pm0.5$ mV [Figs.~1(c-f)], consistent with an induced superconducting gap in the nanowire significantly smaller than the $\sim$3 meV bulk gap of NbTiN we observed.
The zero-field subgap conductance was only a few times smaller than that on the coherence peaks \cite{Takei:2012vj}, and several subgap resonances are evident, most clearly in Fig.~1(d).

Figures 1(c-f) show zero-bias conductance peaks emerging at finite $B_y$ for two settings of $V_g$.
The peaks reached a maximum height of $0.2e^2/h$ above the finite-bias background at $B_y\sim0.5$ T for the gate voltage setting shown in Figs.~1(c, d), and $0.4e^2/h$ for the configuration of Figs.~1(e, f) at the same magnetic field.
The zero-bias peaks persisted up to at least $B_y=1$ T, but above 0.5 T a pair of additional resonances split away from zero bias for the data shown in Fig.~1(c, d).
The peaks for the two configurations both have full widths at half maximum of 0.11 meV, three times larger than expected for thermal broadening based on Coulomb blockade thermometry of device 1 (0.1 K electron temperature).
In the simplest Majorana picture, a non-thermally broadened zero-bias peak should have a quantized height of 2$e^2/h$ \cite{Law:2009gf}, but the zero-bias peak height is expected to be suppressed in short nanowires \cite{Lin:2012fu}.

Zero-bias peaks at finite field, similar to those in Figs.~1(c,d), were also observed for fields oriented perpendicular to the nanowire axis, $B_x$ and $B_z$. 
We did not find a field direction where zero bias peaks were consistently absent for all field strengths. 
This is inconsistent with a simple Majorana picture or perhaps reflects a nonuniform spin-orbit field or irregular Zeeman field resulting from the superconductor covering the wire. Within an antilocalization picture, this observation is consistent with a symmetry class D nanowire (diameter larger than spin-orbit length), but inconsistent with the expected class BDI (diameter smaller than spin-orbit length) \cite{Pikulin:2012fs,Tewari:2012wz,Tewari:2012wd}.

%:Fig 2
%-------------------%
\begin{figure}[t]
\center \label{figure2}
\includegraphics[width=3.4in]{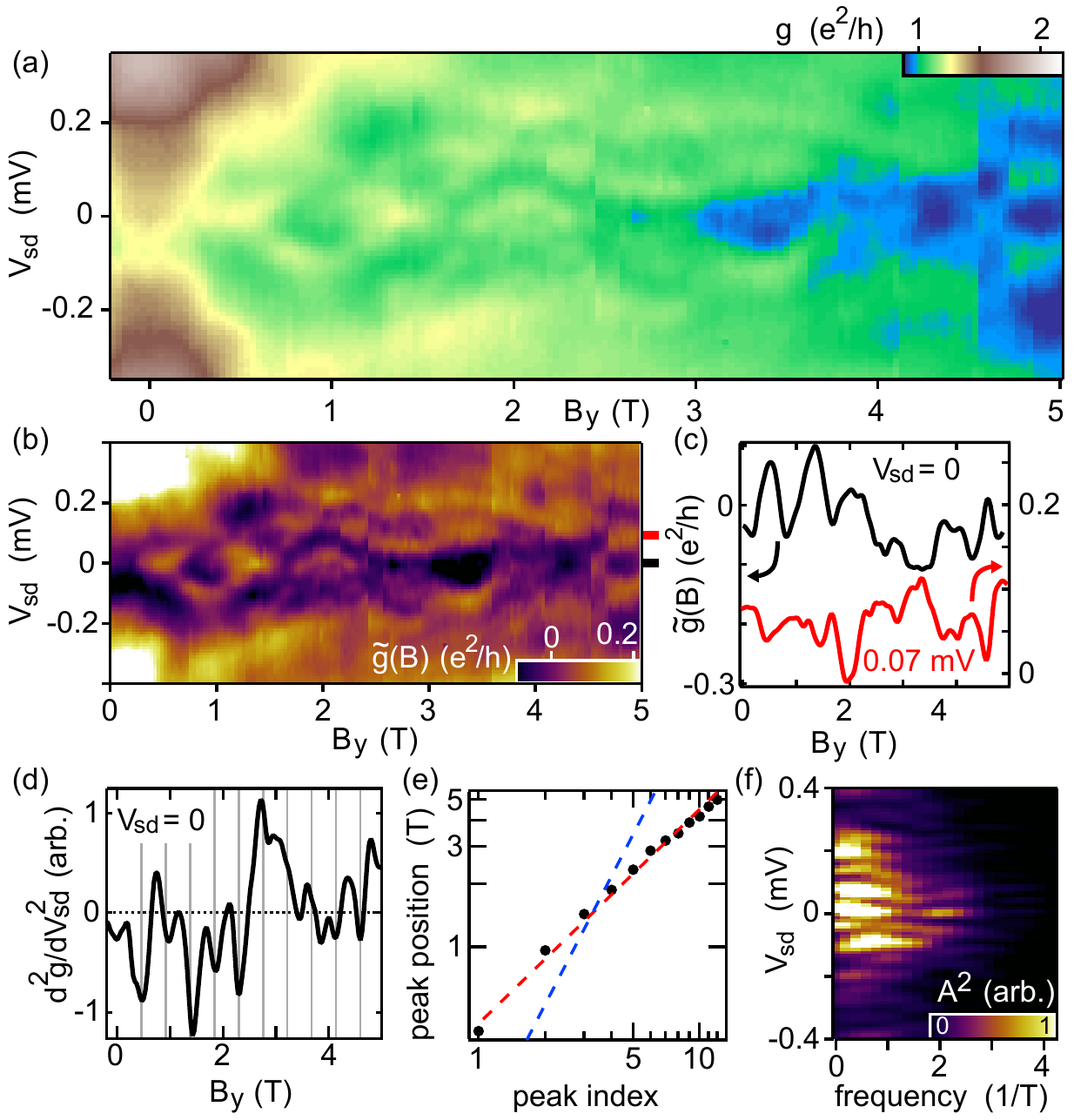}
\caption{\footnotesize{(a) Differential conductance, $g$, as a function of source-drain voltage, $V_{\rm sd}$, and axial magnetic field, \by~(device 1, temperature 0.1 K, $V_g=2.5$ V).  (b) Differential conductance with average over $V_{\rm sd}$ subtracted at each field, $\tilde{g}(B)$. (c) Cuts of the data in (b) at $V_{\rm sd}=0$ (black) and $0.07$ mV (red).  (d) Conductance concavity, $d^2g/dV^2_{\rm sd}$ at $V_{\rm sd}=0$.  Vertical lines are placed at integer multiples of the position of the first large dip in $d^2g/dV^2_{\rm sd}$ at $B_y=0.45$~T.  (e) Peak positions obtained from the local minima in (d) versus peak index.  The red (blue) dashed line has a slope of 1 (2) in the log-log plot.  (f) Squared Fourier amplitudes of $d^2g/dV^2_{\rm sd}$ as a function of \by, calculated row by row at each value of $V_{\rm sd}$. }}
\end{figure}
%-------------------%

The presence of a zero-bias peak did not depend on voltages on the gates under the superconductor within the operating range, $\pm4$ V, though the peak height could be made to vary by up to a factor of 2 using these gates.
Within a Majorana picture, this suggests that the zero-energy end states are either centered beyond the end of the superconductor or that the density under the superconductor cannot be tuned.

%:Fig 3
%-------------------%
\begin{figure}[ht!]
\center \label{figure3}
\includegraphics[width=2.5in]{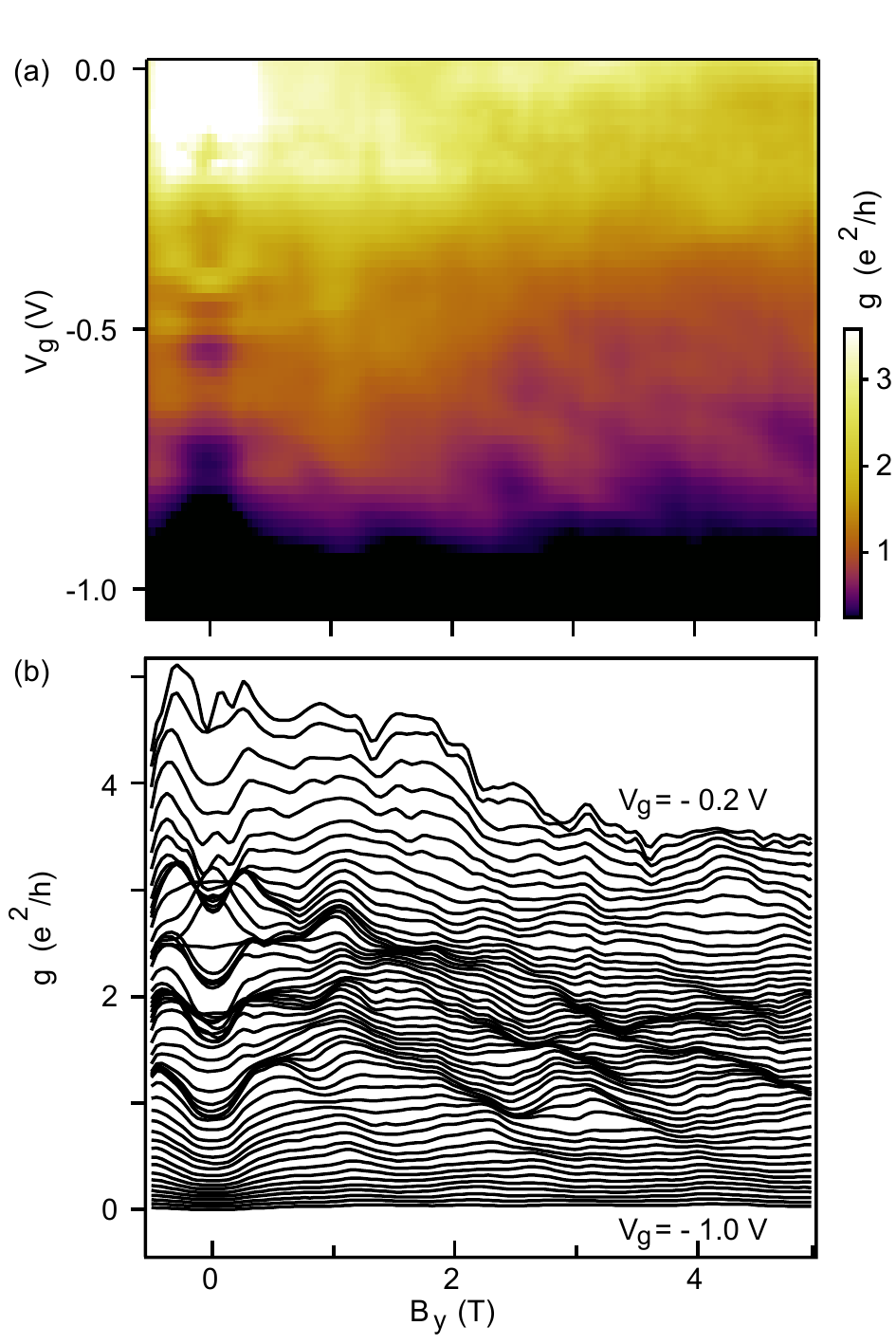}
\caption{\footnotesize{Zero-bias conductance, $g$, as a function of $V_g$ and \by~shown (a) in colorscale and (b) as offset line traces for $V_g$ from -0.2 to -1.0 V.  Data from device 2, temperature 0.5 K.}}
\end{figure}
%-------------------%

At larger values of \by, zero-bias features in both devices showed oscillations in both amplitude and bias position. 
Representative data from device 1 are shown in Fig.~2: As a function of $V_{\rm sd}$ and \by, a zero-bias feature appears at $B_y\sim0.5$~T, splits into two peaks separated by $\sim$0.2 meV, and rejoins into a single zero-bias peak at $\sim$$B_y=1.4$~T [Fig.~2(a)]. At $B_y=1$~T, a faint third peak at zero bias is also visible between the split peaks.
The evolution of these features with larger $B_y$ shows subsequent splitting and rejoining (most clearly for $B_y = 4$-5 T).
A changing background conductance with $B_y$ partially obscures the oscillations.
To highlight these features, we subtract the differential conductance $g$ averaged over $V_{\rm sd}$ at each $B_y$, which defines the quantity $\tilde{g}(B)$ in Fig.~2(b). 
Around 2 T and 4-5 T, the added visibility also reveals peaks splitting and rejoining without the appearance of a third peak at zero bias.
Line cuts of $\tilde{g}(B)$ at $V_{\rm sd}=0$ (black) and 0.07 mV (red) are anti-correlated [Fig.~2(c)]:  peaks at zero bias correspond to dips at 0.07 mV and vice-versa, reflecting checkerboard or criss-cross features in the $B_y$-$V_{\rm sd}$ plane.

Figure 2(d) shows conductance concavity, $d^2g/dV^2_{\rm sd}$, at zero bias as a function of field, highlighting local maxima (negative concavity) and minima (positive concavity) of $g$. 
The concavity appears roughly periodic, with a period equal to the magnetic field at which the first zero-bias peak appears, $B_y=0.45$~T.
Vertical lines in Fig.~2(d) at multiples of 0.45 T approximately line up with the positions of concavity minima (conductance peaks), with two exceptions between 3 and 4 T.
Figure 2(e) shows that the field values of conductance peaks, identified from concavity minima [Fig.~2(d)], increase linearly with magnetic field [Fig.~2(e)].
This periodicity is further reflected in the Fourier transform of concavity as a function of magnetic field at each bias [Fig.~2(f)], indicating a period $\sim 0.5$~T, in addition to a large low-frequency component due to the strongly convex region of $g$ around $B_y=3$~T.
As a function of bias, the Fourier amplitudes are striped, with the $\sim$0.5 T period reappearing at $\pm0.07$ meV.

Oscillatory features were also evident in the zero-bias conductance of device 2 as a function of $V_g$ and $B_y$ (Fig.~3).
Focusing on the gate voltage range between $V_g=-0.5$ and $-1$ V, we observed several approximately parallel diagonal peaks in $g$ with a negative slope in the $B_y$-$V_g$ plane [Fig.~3(a)].
These oscillations were found near pinch-off of the QPC and were not seen for $g>e^2/h$.

Looking over a broader range of gate voltages and fields reveals that the field scale on which the first zero-bias peak appears, \by$\sim 0.5$~T, marks a crossover in conductance that follows opposite trends in the low conductance (tunneling) and high conductance (multichannel) regimes [Fig.~4(a)]. 
In the multichannel regime ($g\gtrsim 2e^2/h$), fields above 0.5~T reduce conductance, while in the tunneling regime, fields above 0.5~T increase conductance. 
These opposite effects of field are qualitatively consistent with a field-driven crossover from a trivial to a topological regime \cite{Wimmer:2011vl}.  Qualitatively similar behavior occurred as a function of $B_x$ and $B_z$.
The dependence of $g$ on $B_y$ is summarized by averaging $g$ versus $V_g$ from Fig.~4(a) over three distinct regions of $B_y$:  $-0.2$ to 0.2 T [blue curve in Fig.~4(b)], 0.5 to 2 T (red curve), and 2.5 to 5 T (black curve). Suppression of $g$ with magnetic field at higher transmission due to the suppression of Andreev reflection would result in a field-induced suppression that scales with $g$, which is not consistent with the data. 

%:Fig 4
%-------------------%
\begin{figure}[ht!]
\center \label{figure4}
\includegraphics[width=3.25in]{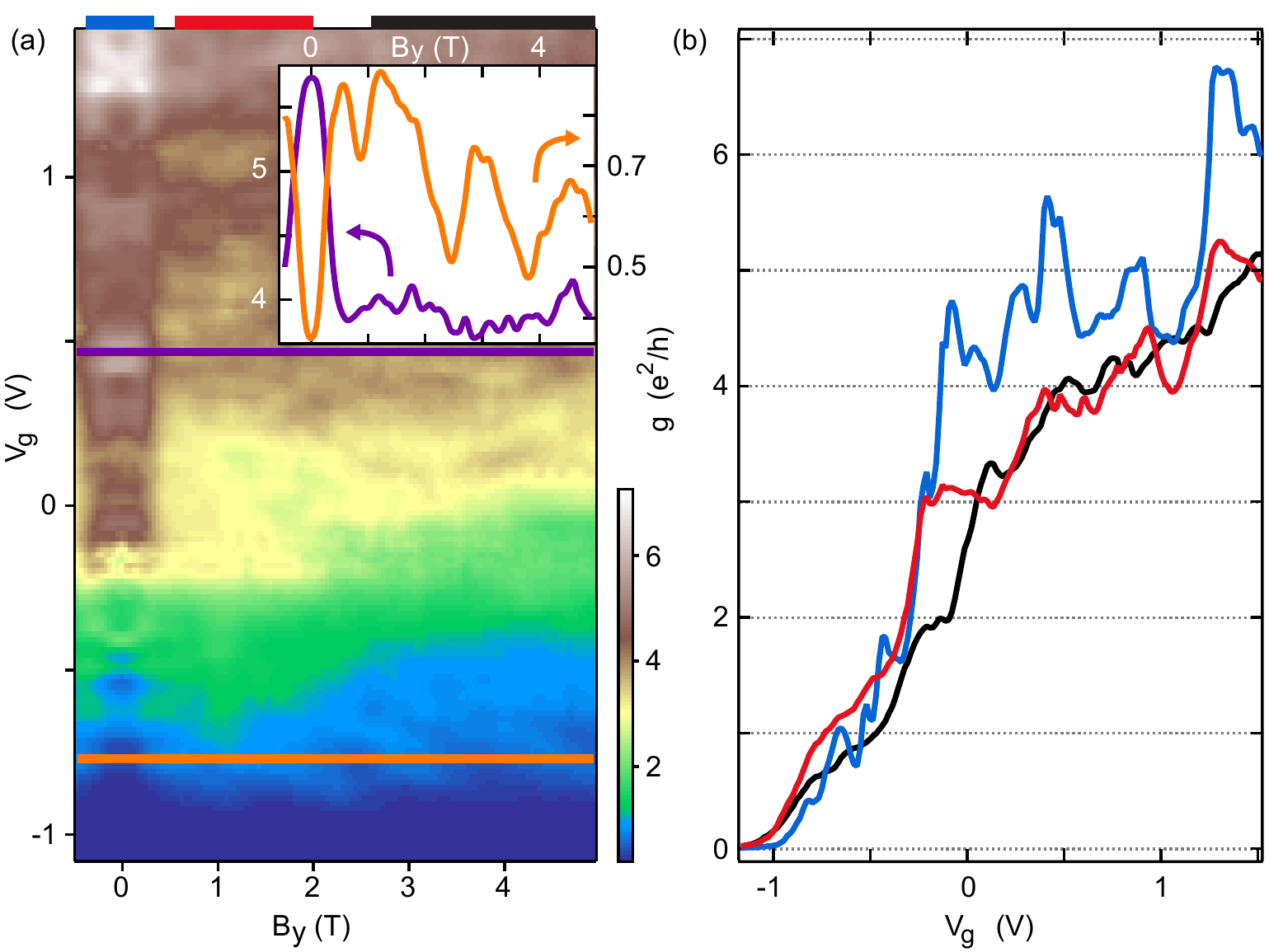}
\caption{\footnotesize{(a) Zero-bias conductance, $g$, as a function of $V_g$ and \by~(device 2, temperature 0.5 K).  Inset:  cuts along \by~at $V_g=-0.75$ V (orange) and $V_g=0.49$ V (purple). (b) $g$ as a function of $V_g$ from (a) averaged over the ranges of \by~indicated by colored bars along the top axis of (a): \by$=-0.2$ to 0.2 T (blue curve), \by$=0.5$ to 2 T (red curve), and \by$=2.5$ to 5 T (black curve).}}
\end{figure}
%-------------------%

The reduction of conductance in the multichannel regime persisted upon raising the sample temperature to 4 K (Fig.~5), still well below the 12 K transition temperature of the NbTiN  leads.
Though smaller, this effect is similar to that at low temperature shown in Fig.~4.
In contrast, elevated temperature did eliminate the zero-bias peak and oscillatory structure. 
We conclude that the three phenomena observed near pinch-off---zero-bias features, enhancement of $g$ with \by, and oscillations---are all associated with an energy scale smaller than the basic phenomenon of induced superconductivity in the nanowire.

Turning now to interpretation, we note that in a short nanowire, separated Majorana zero modes should interact and split away from zero energy.
For a field- and density-dependent Fermi wavelength, interference between zero modes would cause them to split and rejoin as a function of magnetic field and gate voltage \cite{Cheng:2009tw,Prada:2012iv,Rainis:2012wn,DasSarma:2012kt}.
In the simplest picture, peak position in magnetic field is expected to grow with $B^2$ through a combination of a quadratic dispersion and a linear Zeeman effect \cite{Rainis:2012wn}, in contrast with the linear trend we observed.
This picture also predicts an increasing amplitude in $V_{\rm sd}$ of the oscillations with magnetic field, rather than the roughly constant amplitude between 1 and 5 T shown in Fig.~2(a).
The striped pattern evident in Fig.~3 is qualitatively consistent with numerics \cite{DasSarma:2012kt}, though quantitative comparisons are not yet possible.

%:Fig 5
%-------------------%
\begin{figure}[ht!]
\center \label{figure5}
\includegraphics[width=3.4in]{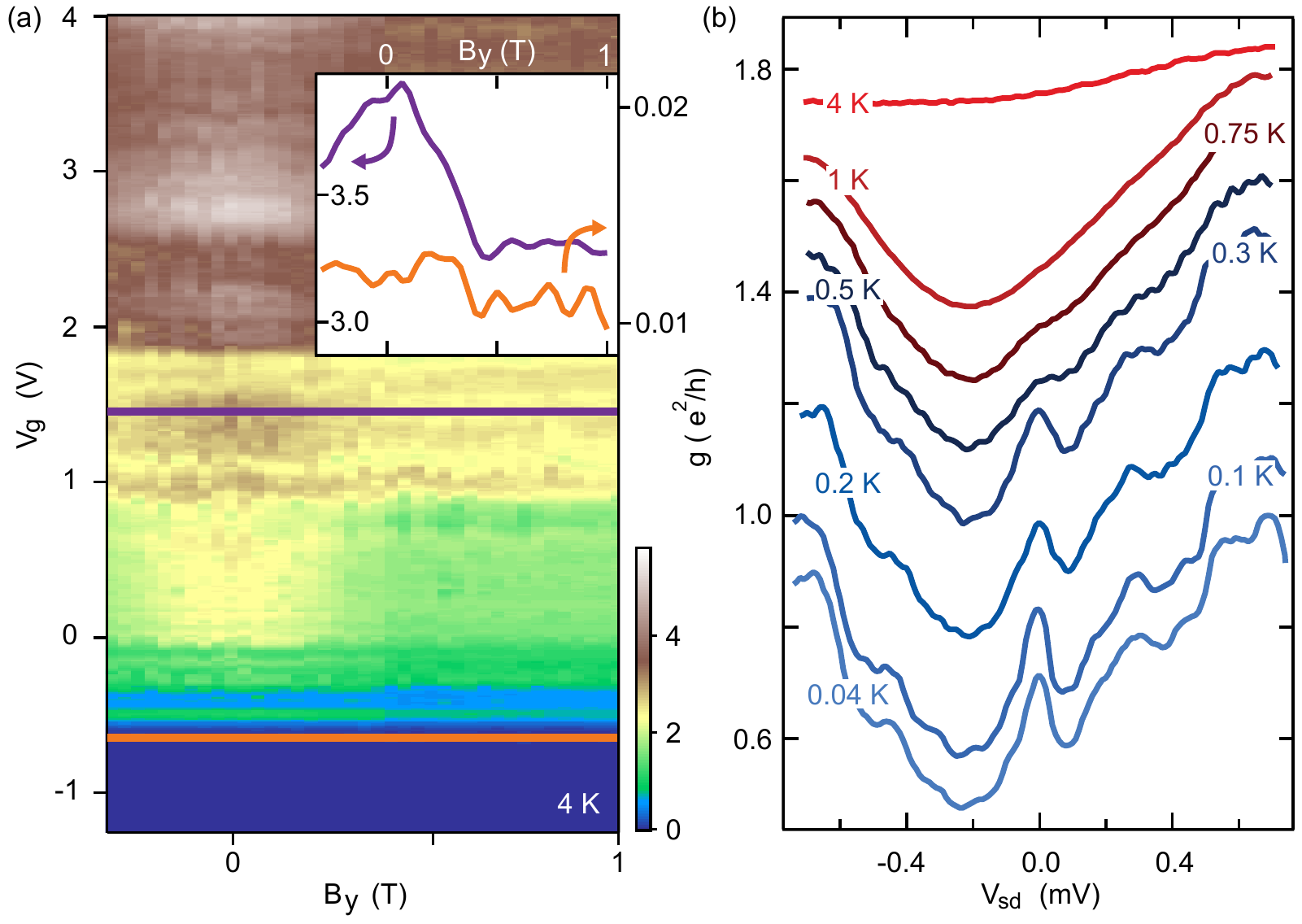}
\caption{\footnotesize{Zero-bias conductance, $g$, of device 2 as a function of $V_g$ and \by~at 4 K.  Inset:  cuts along \by~at $V_g=-0.75$ V (orange) and $V_g=1.49$ V (purple).  (b) $g$ as a function of $V_{\rm sd}$ for device 1 at various temperatures between 0.04 and 4 K at \by = 0.5 T.  Curves are offset vertically for clarity.}}
\end{figure}
%-------------------%

The Kondo effect can also give rise to oscillatory features \cite{DGG:2012ut}.
Oscillatory conductance features could occur as levels that are initially separated by the induced superconducting gap become degenerate or nearly so in a magnetic field.
A similar situation occurs with superconductivity replaced by orbital level spacing in the orbital Kondo effect \cite{JarilloHerrero:2005ck}.
The Kondo effect is usually associated with strong coupling to two normal leads, and a superconducting gap is expected to suppress the effect when the gap is larger than the Kondo temperature.
However, the induced gap in our nanowires is rather ``soft" \cite{Takei:2012vj}, independent of $V_g$.
This soft gap could provide the necessary quasiparticle density of states for Kondo screening \cite{Lee:2012uq,Chang:2012vx,Finck:2012tc}.

The absence of dot-like features in the regime investigated does not rule out the possibility of an interaction-induced spinful state which would then show a zero-bias anomaly via Kondo processes \cite{Thomas:1996wq,Cronenwett:2002ui,Meir:2002tw}.
The oscillations in Fig.~3 persist down to very low conductances ($<10^{-2}\ e^2/h$), consistent with the zero-bias structure in the low-conductance regime of quantum point contacts \cite{Ren:2010vf}.
The constant period of the oscillations [Fig.~2(d)] would require a linear spectrum of subbands above the induced gap, which could result from near-harmonic transverse confinement in the nanowire.
In the limit of strong disorder, the subband spacing is expected to be suppressed completely, resulting in level repulsion set by the spin-orbit energy scale \cite{Liu:2012vj}, roughly in agreement  with the $\sim$0.07 meV scale inferred from the bias dependence of the oscillations in Fig.~2 \cite{NadjPerge:2012wp}.
Within the Kondo interpretation, this energy scale divided by the 0.5 T magnetic field period we observed implies a $g$-factor of 8, significantly smaller than expected for InSb nanowires ($g$-factor $\sim$50) \cite{Nilsson:2009vf} but reasonable when considering repulsion of clustered levels.
The strong temperature dependence of the zero-bias features and oscillations suggests that disorder-induced level crossings without Kondo effect/0.7 structure enhancement do not explain our observations.
Other, less likely origins of the features we observed are discussed in the Supplementary Material.

In the present experiments, differences that clearly discriminate between Majorana zero modes and the Kondo effect/0.7 structure are blurred by disorder and the soft gap. Future experiments must improve these shortcomings or identify other signatures that are robust to them.

We acknowledge discussions with A.~Akhmerov, K.~Burch, W.~Chang, K.~Flensberg, L.~Kouwenhoven, D.~Loss, V.~Manucharyan, S.~Das Sarma, and M.~Wimmer.  Research supported by Microsoft Project Q. The Center for Quantum Devices is supported by the Danish National Research Foundation.
\bibliography{QPC-Refs}

\begin{thebibliography}{46}
\bibitem[{Gno()}]{Gnote}
\bibinfo{note}{$g$ was corrected for 8.4 k$\Omega$ series resistance for both
  devices and 650 $\Omega$ S-contact resistance from 3-wire measurement for
  device 1.}
\end{thebibliography}

\end{document}